\documentclass{article}

\usepackage{iclr2018_conference,times}

\iclrfinalcopy

\usepackage[utf8]{inputenc}
\usepackage[T1]{fontenc}
\usepackage{hyperref}
\usepackage{graphicx}
\usepackage{overpic}
\usepackage{url}
\usepackage{mathrsfs}
\usepackage{subfig}
\usepackage{booktabs}
\usepackage{listings}

\usepackage{bm}
\usepackage{amsmath, amsfonts, amssymb}
\DeclareMathOperator{\sign}{sign}

\DeclareMathOperator{\xnor}{XNOR}
\DeclareMathOperator{\popc}{bitcount}

\usepackage{tikz}
\usepackage{tikzscale}
\usetikzlibrary{calc, positioning, decorations.shapes, matrix}
\tikzstyle{dotted}=[dash pattern=on .05mm off 1.5mm, line cap=round]
\tikzstyle{layer}=[draw, rectangle,%
minimum height=3cm, minimum width=5mm,%
]

\usepackage{pgfplots}
\usepackage{pgfplotstable}
\pgfplotsset{%
  every tick label/.append style={font=\small},
  legend style={draw=none, at={(0.5,1.01)},font=\footnotesize,
    legend columns=-1, anchor=south},
    /pgfplots/ybar legend/.style={
      /pgfplots/legend image code/.code={
        \draw[##1,/tikz/.cd,bar width=3pt,yshift=-0.2em,bar shift=0pt]
        plot coordinates {(0cm,0.8em)};},
    }
  }

\usepackage{xcolor}
\definecolor{turquoise}{cmyk}{0.65,0,0.1,0.3}
\definecolor{purple}{rgb}{0.65,0,0.65}
\definecolor{dark_green}{rgb}{0, 0.5, 0}
\definecolor{orange}{rgb}{0.8, 0.6, 0.2}
\definecolor{red}{rgb}{0.8, 0.2, 0.2}
\definecolor{blueish}{rgb}{0.0, 0.7, 1}
\definecolor{light_gray}{rgb}{0.7, 0.7, .7}
\definecolor{pink}{rgb}{1, 0, 1}

\definecolor{bnet}{RGB}{77,175,74}
\definecolor{neon}{RGB}{55,126,184}
\definecolor{CPU}{RGB}{152,78,163}
\definecolor{GPU}{RGB}{255,127,0}
\definecolor{GPU*}{RGB}{228,26,28}
\definecolor{GPU*64}{RGB}{228,26,28}
\definecolor{GPU*32}{RGB}{228,26,28}

\usepackage[binary-units=true]{siunitx}
\newcommand{\MB}[1]{\SI{#1}{\mega\byte}}
\newcommand{\ms}[1]{\SI{#1}{\milli\second}}
\newcommand{\GBS}[1]{\SI{#1}{\giga\byte\per\second}}

\newcommand{\Figure}[1]{Figure~\ref{fig:#1}}
\newcommand{\Table}[1]{Table~\ref{tab:#1}}

\newcommand{\Section}[1]{Section~\ref{sec:#1}}

\newcommand{\espresso}{\textit{Espresso}}

\newcommand{\gpustar}{\ensuremath{\textsl{GPU}^{\,opt}}}
\newcommand{\cpu}{\textsl{CPU}}
\newcommand{\gpu}{\textsl{GPU}}
\newcommand{\bin}[1]{\ensuremath{#1^b}}
\newcommand{\speedup}[1]{\ensuremath{#1 \times}}

\title{Espresso: Efficient Forward Propagation for \\
Binary Deep Neural Networks}
\author{Fabrizio Pedersoli \\
  University of Victoria \\
  \texttt{fpeder@uvic.ca}
  \And
  George Tzanetakis \\
  University of Victoria \\
  \texttt{gtzan@uvic.ca}
  \And
  Andrea Tagliasacchi \\
  University of Victoria \\
  \texttt{ataiya@uvic.ca}
  \AND
}

\begin{document}

\maketitle

\begin{abstract}
  There are many applications scenarios for which the computational
performance and memory footprint of the prediction phase of Deep
Neural Networks (DNNs) need to be optimized. Binary Deep Neural
Networks (BDNNs) have been shown to be an effective way of achieving
this objective. In this paper, we show how Convolutional Neural
Networks (CNNs) can be implemented using binary
representations. \textit{Espresso} is a compact, yet powerful library
written in C/CUDA that features all the functionalities required for
the forward propagation of CNNs, in a binary file less than 400KB,
without any external dependencies. Although it is mainly designed to
take advantage of massive GPU parallelism, \espresso{} also provides
an equivalent CPU implementation for CNNs. \espresso{} provides
special convolutional and dense layers for BCNNs, leveraging
\textit{bit-packing} and \textit{bit-wise} computations for efficient
execution. These techniques provide a speed-up of
matrix-multiplication routines, and at the same time, reduce memory
usage when storing parameters and activations. We experimentally show
that \espresso{} is significantly faster than existing implementations
of optimized binary neural networks ($\approx$ 2 orders of
magnitude). \espresso{} is released under the Apache 2.0 license and
is available at \url{http://github.com/fpeder/espresso}.
\end{abstract}

\section{Introduction}
\label{sec:intro}

Convolutional Neural Networks have revolutionized computer vision,
pushing the task of object recognition beyond human capabilities
~\citep{krizhevsky2012imagenet,simonyan2014very,szegedy2015going}. Deep
Neural Networks (DNN), have also been successfully applied in other
fields, such as speech
recognition~\citep{graves2013speech,hinton2012deep} and automated
translation~\citep{bahdanau2014neural,sutskever2014sequence}.
Despite achieving impressive classification accuracy results, DNNs
require too much memory and power to be used effectively on embedded
or low-power devices. Many networks consume a considerable amount of
memory.  Memory remains a very limited resource on mobile platforms
making harder the usage of trained DNNs~\footnote{for example, the
  popular AlexNet~\citep{krizhevsky2012imagenet} and
  VGG~\citep{simonyan2014very} architectures consume respectively
  $\approx$ \MB{250} and $\approx$ \MB{520}}. Even when memory is not
an issue, DNNs remain very computationally intensive, and can quickly
drain the battery. Reducing the computational load does not only
improve energy efficiency, but can also enable further
applications. For example, when processing real-time object
classification on mobile, being able to perform faster predictions
frees up computational resources that can be spent on tasks such as
speech recognition and analysis. Therefore, there is a substantial
interest in reducing the computational and memory requirements of
DNNs.

\paragraph{Efficient deep neural networks}
One way to achieve this target is to use specialized hardware for
DNNs. Another strategy is to reduce the network's memory footprint and
associated computation, hence increasing its efficiency. Such
solutions are preferable as they can be implemented in software
without requiring specialized hardware. In our research we follow the
software approach, and focus our attention to \emph{quantized}
networks. In this case, the parameters are stored as ``small''
integers (typically less than $8$-bit) instead of single precision
floating point numbers ($32$-bit).
In particular, we consider the \emph{binary} deep neural networks
(BDNN) proposed by \cite{hubara2016binarized} where parameters and
activations are $1$-bit integers: $\{-1,+1\}$. At the expense of a
relatively small decrease in accuracy, BDNNs can considerably reduce
memory usage, and result in faster execution time (i.e. forward
propagation). Further, note that potential hardware implementation of
BDNNs would also be cheaper due to the reduced number of required
FPUs.
While these results are highly promising, currently only
\emph{proof-of-concept} implementations of BinaryNets have been
published~\citep{hubara2016binarized}.
%
%
Therefore, the availability of a flexible end-to-end framework, with
particular emphasis placed on computational efficiency, can enable
further research on BDNNs, as well as its application to practical
scenarios.

\paragraph{Contributions}
%
%
With \emph{Espresso} we provide an optimized framework for BDNNs
capable of achieving state-of-the-art run-time \emph{performance} with
minimal \emph{memory} footprint
while being numerical equivalent to their non-optimized binary
counterpart.
%
%
\espresso{} provides a complete optimized framework for BDNNs
supporting both the \emph{dense} and the \emph{convolutional} layer.
Current state-of-the-art optimized BDNNs implementations are limited
to the fully connected layer, with the serious drawback of not being
able to run optimized state-of-art convolutional BDNNs (BCNNs).
%
%
%
%
%
While our work is a necessary stepping stone towards optimization of
training routines, in this paper we focus on the optimization of
forward-propagation (i.e. testing), rather than back-propagation
(i.e. training).  \espresso{} is designed to have no external
dependencies. This not only results in a highly optimized
implementation of BDNNs, but also substantially simplifies its
deployment in practical applications, such as those executing on
mobile or embedded devices.

\section{Related Work}
\label{sec:related}

Improving the performance of DNNs can be achieved at either the
hardware or software level. At the hardware level, chipsets that are
dedicated to DNN execution can outperform general-purpose
CPUs/GPUs~\citep{google_tpu,han2016eie}. %
At software level one approach is to design simpler architectures, in
terms of overall floating point operations, that can offer the same
accuracy as the original model~\citep{iandola2016squeezenet}. Another
approach is to prune the weights \citep{guo2016dynamic}, or even
entire filters \citep{li2016pruning}, that have low impact on the
activations such that a simpler model can be derived. These simplified
models can be further compressed by weight sharing
\citep{han2015deep}. Finally instead of removing connections, another
approach is to \emph{quantize} the network weights
\citep{courbariaux2014training} such that computations can be executed
more efficiently.
%

\paragraph{Quantized networks}

In quantized networks, the objective is to train DNNs whose
(quantized) weights do not significantly impact the network's
classification accuracy. For example, \cite{courbariaux2014training}
show that $10$-bits are enough for Maxout Networks, and how more
efficient multiplications can be performed with fixed-point
arithmetic. Continuing this trend, more aggressive quantization
schemes, up to ternary \citep{zhu2016trained}, have also been studied.
%

\paragraph{Binary Deep Neural Networks (BDNN)}

Recently, \cite{courbariaux2015binaryconnect} showed that a network
with \emph{binary} {\small $\{-1,+1\}$} weights can achieve near
state-of-the-art results on several standard datasets. Binary DNNs
(BDNNs) were shown to perform effectively on datasets with relatively
small images, such as the permutation-invariant MNIST~\citep{mnist},
CIFAR-10~\citep{cifar} and SVHN~\citep{svhn}.
Recently, \cite{rastegari2016xnor} show that binarized CNNs can
perform well even on massive datasets such as
ImageNet~\citep{deng2009imagenet} using binarized versions of
well-known DNN architectures such as
AlexNet~\citep{krizhevsky2012imagenet}, ResNet-18~\citep{he2016deep},
and GoogLenet~\citep{szegedy2015going}.  Similarly interesting results
can be achieved by binarizing both DNN weights and activations as
showed by \cite{hubara2016binarized}. In this work, the authors
introduce \textit{BinaryNet}, a technique to effectively train DNNs
where both weights and activations are constrained to {\small
  $\{-1,+1\}$}.
\emph{BinaryNet} achieves nearly state-of-the-art accuracy for MLP
training on MNIST and CNN training on CIFAR-10. The authors also
propose a binary optimized implementation of matrix multiplication
which result in $7\times$ faster performance than the base-line non
optimized implementation, and, almost $2\times$ faster than
Theano~\citep{bergstra2010theano}. Their core contributions, namely to
replace Floating-point Multiply and Add operations (FMAs) with
\emph{XNORs} and \emph{bit-counts}, represent the cornerstone over
which we build our research.

\section{The Espresso Framework}
\label{sec:highlights}
\espresso{} provides the user with the necessary tools for executing
forward-propagation of DNNs, with particular emphasis placed on
convolutional neural networks due to their ubiquitousness in computer
vision applications.
As the complexity of these networks is cubic to the size of the
problem, they are less memory efficient and more computationally
intensive than traditional machine-learning algorithms.
Identifying the memory and computational bottlenecks of DNNs is
therefore essential to enable their practical application.
In particular, our primary focus is \emph{GPU-optimized} BDNN
architectures, which we refer to as~\gpustar, but we also support the
equivalent floating-point counterparts on heterogeneous architectures,
which we refer to as~\cpu{} and~\gpu{}. The \cpu{} and \gpu{}
implementations of \espresso{} do not feature binary optimizations
because the data is encoded as single precision floating point
numbers. However they still utilize an optimized library for matrix
multiplication.

\paragraph{Hybrid DNNs}

The \espresso{}'s implementations of tensors and
layers come in three variants $\{\cpu,\gpu,\gpustar\}$. A CPU-tensor
is allocated in CPU memory, and is processed on the CPU using
sequential code. A GPU-tensor is allocated on GPU main memory and is
processed by CUDA kernels. \espresso{} provides functions for
\emph{converting} tensors and layers from one variant to the other,
and different variants can also be interconnected with each
other. Consequently, \espresso{} enables the design of hybrid DNNs
consisting of a combination of $\{\cpu,\gpu,\gpustar\}$ layers.

\paragraph{The computational bottleneck: dot products}

Dense linear algebra is at the heart of deep-learning as deep networks
can be viewed as a composition of \emph{matrix-matrix},
\emph{matrix-vector} and \emph{elementwise matrix-matrix or
  vector-vector} multiplications. The implementation of these dense
linear algebra operations relies heavily on the efficient computation
of the \textit{dot-product}. The execution of this operator consists
of (single precision) \emph{Floating-point Multiply and Add} (FMA)
operations.  In modern architectures, floating-point multiplications
executing on the FPU dominate the complexity of FMAs, and BDNNs
address these concerns by replacing FMAs with simpler \emph{bitwise}
operations; see~\Section{bin-net}.

\paragraph{Technical highlights}

The superior computational performance of \espresso{} derives from
three main technical contributions: (1) the use of bit-packing in
network layers, (2) better memory layout and management, and (3) the
use of custom optimized CUDA kernels. Through the use of bit-packed
layers, \espresso{} can execute a forward operation without the need
for expensive memory re-arrangements employed by existing
implementations. As dynamic memory allocation on GPUs is a performance
bottleneck, \espresso{} implements a custom memory allocator that
  pre-allocates memory at start-up, and replaces the traditional
  \emph{malloc} and \emph{free} system calls.
%
Finally, matrix multiplications are performed with CUDA kernels that
have been adapted to bit-packing, and only resort to XNORs and bit-counts.

\section{Binary Deep Neural Networks (BDNNs)}
\label{sec:bin-net}

In this section, we overview the fundamental characteristics of
BDNNs~\citep{hubara2016binarized} that inform the basics of
\espresso{}'s design. In BDNNs, computationally intensive FMA
operations are replaced by \textit{XNOR} (for multiplications) and
\textit{bit-count} (for additions), enabling significant computational
speed-ups. In particular, XNOR is a simpler machine instruction
compared to floating point multiplication, and therefore achieves much
higher throughput on many architectures. More importantly, a single
XNOR step can execute multiple $64$-bit wide blocks of dot-products,
further increasing the overall computational efficiency.  In what
follows, we describe how a network is binarized, detail a compressed
memory layout enabling efficient execution of dot-products, show how
to re-interpret input data to allow execution on fixed-precision input
(e.g. images), and provide a few notes regarding the training
procedure.

\subsection{Network binarization}

A BDNN is composed of a sequence of $k=1,\ldots,L$ layers whose
weights $\bin{W}_k$ and activations $\bin{a}_k$ are binarized to the
values $\{-1,+1\}$. The superscript $b$ in the notation indicates
binary quantities. Weights and activations are $\{-1,+1\}$, but at the
hardware level they must be encoded as $\{0,1\}$. Our convention is to
encode $-1\rightarrow 0$ and $+1\rightarrow 1$. Amongst many possible
choices, e.g. stochastic
binarization~\citep{courbariaux2015binaryconnect}, we employ the
following activation function due to its efficient implementation:
\begin{equation}
  \bin{x} = \sign(x) =
  \begin{cases}
    +1 \quad x\geq0 \\
    -1 \quad \text{otherwise}
  \end{cases}
\end{equation}

\subsection{Bit-packing}

The weights of a BDNN can be stored in the bits of a $64$-bit
word. One immediate advantage of bit-packing is to drastically reduce
the memory usage by a $32\times$ factor. An even more significant
advantage is the ability to process multiple values at the same time
using registers. This is particularly useful for dot-products: with
bit-packing we can compute a dot-product of $64$ element vectors by
using just one XNOR and one bit-count. Furthermore, modern computer
architectures provide a hardware instruction for counting the number
of bits set to 1 in a given word. Assuming binary vectors
$a, b\in\mathbb{B}^{1\times N}$ where $N$ is a multiple of $64$, the
dot-product is then equivalent to:
\begin{equation}
  a\cdot b \;\equiv\; N - \left( \sum_{i=1}^{N/64}\popc(\xnor(a_i,b_i)) \right)
  \ll 1 \;\triangleq\; a\odot b
\end{equation}
where $\ll$ represents the bit-shift operator. This simple computation
becomes the building block of optimized BDNNs as binary matrix-matrix
or matrix-vector operations are computed in this fashion.

\subsection{Input data binarization}

BDNNs require binary input data, which is not typically available at
the first layer of the network. However, the input data usually comes
in a fixed precision format (e.g. $8$-bit/channel in RGB
images). Therefore, the optimized computation of dot-products can
still be applied if we split the input data according to bit-planes,
and then sum back each contribution according to the corresponding
weight. For instance, if with $\langle a\rangle_n$ we indicate the
$n$-th bit of a fixed precision vector, and with $i$ the corresponding
bit-plane, we obtain:
\begin{equation}
  a\cdot b \equiv \sum_{i=0}^{n-1} 2^i \langle a \odot b\rangle_i
\end{equation}

\subsection{Training}

When training a BDNN, it is important to note that the gradient is
computed with the binary weights, but is accumulated with floating
point precision~\citep{hubara2016binarized}. That is because the
optimizer needs sufficient precision to make a reliable update. In
addition, the derivative of the sign function, which is zero almost
everywhere, cannot be used for back-propagation. To overcome these
issues, the \textit{straight-through
  estimator}~\citep{bengio2013estimating} is employed, where $1$ is
back-propagated if the floating point argument $|x|\leq 1$, and $0$
otherwise. Finally, during training weights are clipped to $[-1,1]$ to
avoid a large growth of the floating point weights that would not have
an impact on the binary weights.

\section{Espresso architecture}
\label{sec:architecture}

The principal components of our framework are \emph{tensors},
\emph{layers} and the \emph{network}. These components are
organized as a hierarchy. Tensors are $n$ dimensional matrices used
for storing \emph{inputs}, \emph{weights} and \emph{activations}
(outputs). A layer processes an input tensor and produces an output
tensor, while a network consists of a concatenation of layers.

\subsection{Tensors}

In \espresso{}, each element of a tensor
$A \in\mathbb{R}^{M\times N \times L}$ is identified by the triplet
$m,n,l$, where $m \in [0,M)$ indicates the row, $n\in [0,N)$ indicates
the column, and $l\in [0,L)$ indicates the channel. A tensor is stored
in memory using row-major order with interleaved channels. Therefore,
according to this layout, the element $A_{m,n,l}$ is found at position
$(mN +n)L + l$ in linear memory.
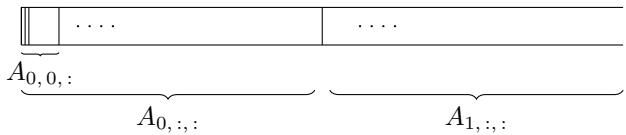
\begin{figure}[htb]
  \centering
  \begin{tikzpicture}[node distance=1mm,inner sep=.2mm]


  \def\x{8}\def\y{.5}\def\dx{.2}\def\dy{.2}
  \coordinate (a) at (0,0);
  \coordinate (b) at (4,0);
  \coordinate (c) at (.5,0);

  \draw (a) -- ++(0,\y) -- ++(\x,0);
  \draw (a) -- ++(\x,0);
  \draw (b) -- ++(0,\y);
  \draw (c) -- ++(0,\y);

  \foreach \k in {0,.05,...,.1} {
    \draw ($(a)+(\k,0)$) -- ++(0,\y);
  }

  \draw[dotted,thick] (4.5,\y/2) -- ++(.5,0);
  \draw[dotted,thick] (.75,\y/2) -- ++(.5,0);

  \draw[decorate,decoration={brace,mirror,amplitude=1.2},yshift=-1mm]
  (0,0) -- (.5,0) node [midway,below,yshift=-1mm]
  {$\displaystyle A_{0,\,0,\,:}$};

  \draw[decorate,decoration={brace,mirror,amplitude=1.2mm},yshift=-6mm]
  (0,0) -- (3.9,0) node [midway,below,yshift=-2mm]
  {$\displaystyle A_{0,\,:,\,:}$};

  \draw[decorate,decoration={brace,mirror,amplitude=1.2mm},yshift=-6mm]
  (4.1,0) -- (8,0) node [midway,below,yshift=-2mm]
  {$\displaystyle A_{1,\,:,\,:}$};

\end{tikzpicture}

  \label{fig:tensor-layout}
\end{figure}

We use the notation $A_{m,n,:}$, to indicate all the channels of the
$(m,n)$-th element. Using the same storing scheme \espresso{} also
defines bit-packed tensors for $\gpustar$ implementations but with the
following changes to further increase its performance.  Bit-packing is
performed according to the number of channels: when $L>1$ bit-packing
is done along the $l$ dimension; when $L=1$ bit-packing is done along
the $n$ dimension.
\begin{figure}[h]
  \centering
  \resizebox{.7\textwidth}{!}{\begin{tikzpicture}
  \node at (-.5,2) {convolutional};
    \def\x{1} \draw[-] (0,0) -- (0,\x) node[midway,left] {$m$} --
    (\x,\x) -- (\x,0) -- (0,0); \def \y{2}
    \draw[-] (0,\x) -- ++(\y,\y); \draw[-] (\x,\x) -- ++(\y,\y);
    \draw[-] (\x,0) -- ++(\y,\y) node[midway,right] {$n$}; \draw[-]
    ($(0,\x)+(\y,\y)$) -- ++(\x,0) node[midway,above] {$l$} --
    ++(0,-\x);

    \def\o{.15}
    \foreach \asd in {0,.2,...,.4} {
      \draw[fill] ($(\asd,\x)+(\o,-\o)$) circle (.3mm);
      \draw[->, gray] ($(\asd,\x)+(\o,-\o)$) -- ++(\y,\y);
    }

    \node at (.5,-.5) {$M\times N\times L$};

    \def\s{2.5}
    \draw[-] (\s,0) -- ++(0,\x) -- ++(\x,0) -- ++(0,-\x) -- ++(-\x,0);
    \def\y{.3}
    \draw[-] (0,\x) ++(\s,0) --  ++(\y,\y);
    \draw[-] (\x,\x) ++(\s,0) -- ++(\y,\y);
    \draw[-] (\x,0) ++(\s,0) --  ++(\y,\y);
    \draw[-] ($(0,\x)+(\y,\y)$) ++(\s,0) -- ++(\x,0) -- ++(0,-\x);

    \node at (\s+.5, -.5) {$M\times N\times \frac{L}{64}$};

    \def\s{5.5}
    \draw[-] (\s,1) -- ++(0,2) node[midway,left] {$m$} --
    ++(3,0) node[midway,above] {$n$} -- ++(0,-2) -- ++(-3,0);

    \foreach \asd in {0, .2, ..., .4} {
      \draw[fill] ($(\s+.2, 2.8-\asd)$) circle (.3mm);
      \draw[->,gray] ($(\s+.2, 2.8-\asd)$) -- ++(2.6,0);
    }

    \node at (\s+1.5,.5) {$M\times N\times 1$};
    \node at (\s+1.5,0) {dense};

    \coordinate (caz) at (9,3);
    \draw[-] (caz) -- ++(1,0) -- ++(0,-2) -- ++(-1,0) -- ++(0,2);

    \node at (9.5,.5) {$M\times \frac{N}{64}\times 1$};
  \end{tikzpicture}

}
  \label{fig:tensor-packing}
\end{figure}
For \emph{convolutional} layers this packing direction enables
efficient memory access when unrolling/lifting a tensor, which would
have not been possible if either $m$ or $n$ had been chosen
instead. More specifically, this layout is optimal for retrieving a
pixel neighborhood as needed by convolution without requiring the
layout to be changed. Further, typically a large number of filters are
used resulting in an increase of tensor dimension in the $l$
direction, while the $m$ and $n$ dimensions are progressively shrunk
by pooling layers.  For other layer types, $n$ is the most efficient
packing direction, as neurons are stored along rows and their number
decreases as we move toward later stages in the network.

\subsection{Layers}

\espresso{} provides the following layer types: \emph{Input},
\emph{Convolutional}, \emph{Pooling}, \emph{Dense} (i.e. fully
connected) and \emph{Batch-normalization}. Each layer is characterized
by its size, tensor parameters and output. The \espresso{} API defines
for each layer a \textit{forward} function that computes the output of
a layer given an input tensor, and a function for applying
\textit{non-linearity} to the outputs of convolutional and dense
layers.  Moreover, the convolutional layer features additional
functions for \emph{pooling} and \emph{unrolling}.

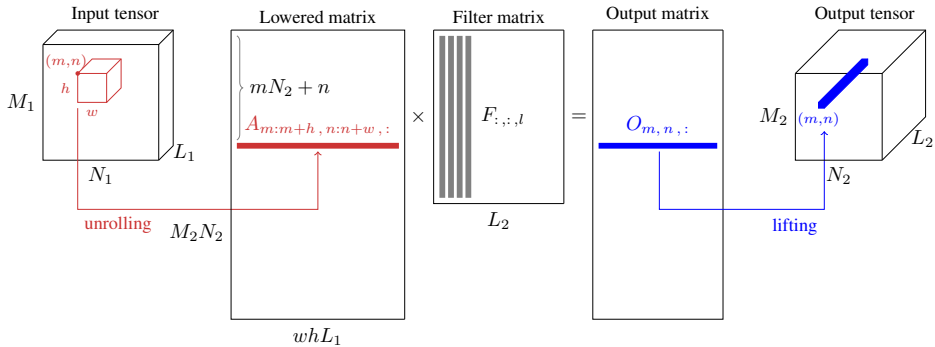
\begin{figure}[htb]
  \centering
  \resizebox{.9\textwidth}{!}{\begin{tikzpicture}
    \def\px{-.5}\def\py{2.25}
    \def\x{2}\def\y{2}\def\l{.25}
    \coordinate (p) at (\px,\py);
    \draw (p) -- ++(\x,0) -- ++(0,-\y);
    \draw (p) -- ++(0,-\y) node[midway,left] {$M_1$} --
    ++(\x,0) node[midway,below] {$N_1$};
    \draw (p) -- ++(\l,\l) --  ++(\x,0)
    node[midway,above] {\small Input tensor}  --
    ++(0,-\y) -- ++(-\l,-\l) node[midway,right] {$L_1$};
    \draw ($(p)+(\x,0)$) -- ++(\l,\l);

    \begin{scope}[every path/.style={red}]
      \def\x{.5}\def\y{.5}
      \coordinate (p) at ($(p)+(.6,-.5)$);
      \node[xshift=-2mm,yshift=.2cm] at (p) {$\scriptstyle (m,n)$};
      \draw[fill] (p) circle (.3mm);
      \draw (p) -- ++(\x,0) -- ++(0,-\y)
      -- ++(-\x,0) node[midway,below] {$\scriptstyle w$}
      -- ++(0,\y) node[midway,left] {$\scriptstyle h$};
      \draw (p) -- ++(\l,\l) -- ++(\x,0) -- ++(0,-\y) -- ++(-\l,-\l);
      \draw ($(p)+(\x,0)$) -- ++(\l,\l);
    \end{scope}

    \draw[red, ->] ($(p)+(0,-.6)$) -- ++(0,-1.75)
    node[right,below,xshift=.7cm] {\small unrolling} -| (4.25,.4);

    \def\px{2.75}\def\py{2.5}
    \def\x{3}\def\y{5}\def\w{.5}\def\s{2}
    \coordinate (a) at (\px,\py);
    \draw (a) -- ++(\x,0)
    node[midway,above] {\small Lowered matrix} -- ++(0,-\y);
    \draw (a) -- ++(0,-\y)
    node[midway,left,yshift=-1cm] {$M_2N_2$} -- ++(\x,0);

    \draw[red,line width=1mm] ($(a)+(.1,-\s)$) -- ++(\x-.2,0)
    node[midway,above] {$\displaystyle A_{m:m+h\,,\,n:n+w\,,\,:}$};
    \node[yshift=-3mm] at (\px+\x/2,\py-\y) {$whL_1$};

    \draw[decorate,thin,gray,decoration={brace,amplitude=1mm}]
    ($(a)+(+1mm,-1mm)$) -- ++(0,-\s+.2)
    node[black,midway,right,xshift=1mm] {$mN_2+n$};

    \def\px{6.25}\def\x{2.25}\def\y{3}
    \coordinate (a) at (\px, \py);
    \draw (a) -- ++(\x,0)
    node[midway,above] {\small Filter matrix};
    \draw (a) -- ++(0,-\y) node[midway,left] {$\times$} -- ++(\x,0);
    \draw ($(a)+(\x,0)$) -- ++(0,-\y) node[midway,right] {$=$};

    \foreach \x in {0,.15,..., .50} {
      \draw[gray,line width=1mm] ($(a)+(.15+\x,-.1)$) -- ++(0,-\y+.2);
    }
    \node at ($(a)+(1.2,-\y/2)$)
    {$\displaystyle F_{:\,,:\,,l}$};

    \node[yshift=-3mm] at (\px+\x/2,\py-\y) {$L_2$};

    \def\px{9}\def\x{2.25}\def\y{5}
    \coordinate (a) at (\px,\py);
    \draw (a) -- ++(\x,0) node[midway,above] {\small Output matrix};
    \draw (a) -- ++(0,-\y) -- ++(\x,0) -- ++(0,\y);

    \draw[blue,line width=1mm] ($(a)+(.1,-\s)$) -- ++(\x-.2,0)
    node[midway,above] {$\displaystyle O_{m,\,n\,,\,:}$};

    \draw[blue,->] ($(a)+(1.15,-\s-.1)$) -- ++(0,-1) -| (13,.75)
    node[midway,left,yshift=-3mm] {\small lifting};

    \def\px{12.5}\def\py{1.75}
    \def\x{1.5}\def\y{1.5}\def\l{.75}
    \coordinate (p) at (\px,\py);
    \draw (p) -- ++(\x,0) -- ++(0,-\y);
    \draw (p) -- ++(0,-\y) node[midway,left] {$M_2$} --
    ++(\x,0) node[midway,below] {$N_2$};
    \draw (p) -- ++(\l,\l) --  ++(\x,0) --
    ++(0,-\y) -- ++(-\l,-\l) node[midway,right] {$L_2$};
    \draw ($(p)+(\x,0)$) -- ++(\l,\l);

    \begin{scope}[every path/.style={blue}]
      \def\x{1mm}\def\y{1mm}
      \coordinate (p) at ($(p)+(.4,-.5)$);
      \node[yshift=-.3cm] at (p) {$\scriptstyle (m,n)$};
      \draw[fill] (p) -- ++(\x,0) -- ++(0,-\y) -- ++(-\x,0) -- ++(0,\y);
      \draw[fill] (p) -- ++(\l,\l) -- ++(\x,0)
      node[black,midway,above,yshift=.5cm] {\small Output tensor}
      -- ++(0,-\y) -- ++(-\l,-\l);
      \draw ($(p)+(\x,0)$) -- ++(\l,\l);
    \end{scope}

  \end{tikzpicture}

}
  \caption{\emph{unrolling} and \emph{lifting} operations for CNN
    layers }
\label{fig:unrolling}
\end{figure}

\paragraph{Convolutional layers}
In our framework, 2D convolutions are computed through matrix
multiplications -- an operation involving a very high reuse of data.
For both \cpu{} and \gpu{}, this computation is performed by
sectioning data in amounts that are
cache-friendly~\citep{dongarra1990set}, resulting in implementations
attaining close to peak computational performance. However, in order
to express convolution as matrix multiplication we need to re-organize
the input memory appropriately.  This is achieved through the
\textit{unrolling} procedure; see \Figure{unrolling}. It consists of
transforming a tensor into a matrix where each row is formed by
unrolling the tensor data contained in each convolution sliding
volume. The unrolled matrix is then multiplied by the filter
matrix. Finally, the result of the convolution is reordered back to
tensor by using the \textit{lifting} procedure. In \espresso{} we do need
to manually lift the convolution result in order to undo the
unrolling: thanks to our tensor representation this happens
automatically and at zero cost. \espresso{} provides CUDA kernels for the
\emph{unrolling} and \emph{pooling} of tensors for both \gpu{} and
\gpustar{} implementations.

\paragraph{Efficient Matrix multiplication}
Matrix-vector multiplications are fundamental operations of both dense
and CNN layers. For the \cpu{} architecture, we use the OpenBLAS
library~\citep{openblas} to implement these operations. For \gpu and
\gpustar architectures, the CUDA kernels are based on
MAGMA(sgemm)~\citep{magma}, modified to make it compatible with
our binary data representation.  These kernels for matrix
multiplication feature \textit{register blocking} optimization: since
the introduction of \emph{Fermi} architectures the number of registers
have been increased, while register access latency has been
substantially reduced compared to shared-memory; hence caching at the
register-memory level results in considerably faster
throughput~\citep{nath2010improved}. \espresso{} first fetches the tiles
of the two matrices into shared-memory and then process sub-tiles
using registers.  In the \gpustar{} variant, we modify the code by
replacing blocks of $64$ (or blocks of $32$ for \gpustar{} $32$) single
precision multiply and add (FMA) operations with XNOR and bit-count
using packed tensors. We also re-tune the kernel block size parameters
for improving the performance on reduced size matrices.

\paragraph{Zero-padding for convolutions} Typical CNN implementations
apply a tensor convolution in a ``same'' configuration, where the
sizes of input and output tensors matches. This is achieved by
zero-padding input tensors, but in convolutional \gpustar{} layers the
\emph{zero-padding} of the input introduces the side-effect of making
the data ternary $\{-1,0,+1\}$. We deal with this problem by treating
the data as if it was binary (zero is considered a minus one) and fix
the results of the convolution at these corner-cases in
post-processing. This allows us to leave the convolution kernel code
-- the computational bottleneck of the code -- untouched. The
corner-cases are fixed using a highly efficient kernel which executes
an element-wise sum between the results of the convolution and the
correction matrix. The correction matrix is computed once, when the
\gpustar{} layer is loaded, and it simply consists of the convolution
of the layer's weights with a {\small $(+1)$}-padded zero-tensor.

\paragraph{Converting a network to Espresso}
A DNN in \espresso{} is defined as a combination of layers, which is
loaded at run-time by reading its parameters file. The parameters file
specifies the storage format of all the layers, as well as their
weights. Therefore, it completely specifies a DNN as layers are stored
sequentially. Training of the network is done by
BinaryNet~\citep{hubara2016binarized}; the resulting parameters are
converted to the \espresso{} format by utility script distributed
together with our sources.

\section{Evaluation}
\label{sec:experiments}

The performance of our framework is evaluated in terms of average
computational time needed to perform a particular task. The execution
times, averaged over $100$ experiments, are obtained on a machine
equipped with an NVIDIA GeForce GTX 960 with $2$GB of RAM, and a
Intel\textregistered{} dual-Xeon\textregistered{} X5660 @ $2.80$
GHz. In \cpu{} mode, we configure the OpenBLAS library for matrix
multiplication to use all the $24$ available cores.

\paragraph{Experimental design}

We perform three quantitative evaluations: (\Section{matmul})~matrix
multiplications of two dense square matrices of size
$8192 \times 8192$; (\Section{mlpnist})~forward-propagations of a
Multi-Layer Perceptron (MLP) trained on the MNIST
dataset~\citep{mnist}; (\Section{cnncifar})~forward-propagations of a
Convolutional Neural Network (CNN) trained on the CIFAR-10
dataset~\citep{cifar}. By Using the same models and datasets, we
compare \espresso{} with: (1)~the author provided optimized
implementation of BinaryNet \citep{courbariaux2015binaryconnect};
(2)~the optimized BDNN implemented in the Intel Nervana \emph{neon}
framework~\citep{neon}; (3)~a self-comparison
across~$\{\cpu{},\gpu{},\gpustar{}\}$~as no binary-optimized
implementations of convolutional layers are publicly
available. \espresso{} is numerically equivalent to BinaryNet in terms
of classification accuracy. Therefore our evaluation focuses on
computation speed.

\paragraph{Public datasets}

The MNIST dataset~\citep{mnist} consists of $60$K instances for
training and, $10$K instances for testing. Each instance is a
$28\times 28$ grayscale image that depicts digits ranging from $0$ to
$9$. The CIFAR-10 dataset~\citep{cifar}, consists of $50$K training
instances and $10$K testing instances of $32\times 32\times 3$ color
images. Images are subdivided into $10$ classes (airplanes,
automobiles, birds, cats, deers, dogs, frogs, horses, ships and
trucks). Since our interest is to asses the real-time performance of
binary optimized DNNs, in those experiment we use a batch-size of one,
and measure the averaged forward time for each image of the
testing-sets for each dataset.

\subsection{Binary dense matrix multiplication}
\label{sec:matmul}

\begin{table}[htbp]
  \centering
  \caption{Averaged time of binary optimized matrix
    multiplication.}
  \label{tab:res-matmul}
  \begin{tabular}{rrr}
    \toprule
    BinaryNet & Espresso \gpustar{} \small{32-bit} & Espresso \gpustar{}\small{64-bit} \\
    \midrule
    \ms{88} & \ms{16} (\speedup{5.5}) & \ms{11} (\speedup{8})\\
    \bottomrule
  \end{tabular}

\end{table}

In computing dense matrix multiplication, \espresso{} outperforms
BinaryNet by a \speedup{\approx 8} factor. Much of the gain can be
attributed to our optimized kernels, and the use of register blocking:
by fetching bigger data from main memory and shared memory, our kernel
increases the bandwidth utilization by decreasing the number of memory
fetch instructions.  The use of $64$-bit packing instead of the
$32$-bit~(such as that of BinaryNet), introduces an additional
performance improvement. The $64$-bit kernel achieves a memory DRAM
throughput of \GBS{40} for reads and \GBS{5} for writes, while the
$32$-bit kernel obtain \GBS{29.6} for reads and \GBS{3.6} for
writes. This translates into the resulting $\approx 25\%$ speed
improvement.

\subsection{Multi-layer perceptron on MNIST}
\label{sec:mlpnist}

\begin{table}[htbp]
  \centering
  \caption{Average prediction time of the BMLP.}
  \label{tab:res-mlp}
  \begin{tabular}{rrrrr}
    \toprule
    BinaryNet & Nervana/Neon & Espresso \cpu{} & Espresso \gpu{} & Espresso \gpustar{}\\
    \midrule
    \ms{18} & \ms{17} & \ms{37.4} & \ms{3.2} (\speedup{5.6}) & \ms{0.26} (\speedup{68})\\
    \bottomrule
  \end{tabular}

\end{table}

We evaluate the average classification execution time over the MNIST
dataset, where we trained the MLP architecture from
\citep[Sec~2.1]{courbariaux2016binarized} with author-provided
sources, and then converted it to \espresso{}'s format. In
\Table{res-mlp}, \espresso{} achieves a consistent speed-up
of~\speedup{\approx 68} when compared to BinaryNet. As the
Nervana/neon implementation of binary network is a BinaryNet
derivative, it is affected by the same drawbacks of BinaryNet, and
hence achieves comparable performance. Both alternatives have the
additional cost of running CUDA through Python/Theano which may
introduce further latency in the process.
In \Table{res-mlp}, the evaluation over the three variants of
\espresso{} shows the expected outcome, with the $\gpustar$
implementation leading the ranking. Note that we are able to achieve a
speedup of \speedup{\approx 12} on an NVIDIA GTX 960 ($\approx 2.5$
TFLOPs), although this device has only roughly four times more
throughput than the Xeon X5660 ($\approx 500$ GFLOPs without
turbo-boost). Through binary optimization, we are able to further
increase the performance to \speedup{\approx 15} with respect to the
GPU implementation.
We attribute our computational gains to~(1) the use of
\emph{binary-optimized} layers,~(2) our use of \emph{optimized
  kernels} for matrix multiplication and~(3) \espresso{}'s ability to
perform binary optimization of the first layer.

\paragraph{Binary optimized layers}
An evident drawback of Binary-Net is the need for binarizing/packing
the layer's parameters \emph{every time} a forward method is
called. In the case of binary optimized networks, the cost of packing
the parameters is closely related to the cost of multiplication
itself. Therefore, the reduction of bit-packing function calls leads
to a consistent improvement. This motivates our choice of designing
specific layers, where bit-packing is done once during network
loading.

\paragraph{Optimized kernels}
BinaryNet employs two bit-packing kernels: one for row-packing, the
other for column-packing. Although BinaryNet's pack-by-rows kernel is
slightly slower than ours ($8\%$), the pack-by-columns kernel is
significantly slower (\speedup{\approx 4}) due to non-coalesced accesses to
global memory. An additional performance gain of $\approx 15\%$ is
achieved by swapping matrix-vector in favour of matrix-matrix
multiplication kernels when appropriate (i.e. Dense layers with batch
size equal to $1$); for this reason, \espresso{} also includes the
binary-optimized MAGMA(sgemv) kernel.

\paragraph{First-layer binary optimization}
Another important advantage offered by \espresso{} is the ability to
leverage binary optimization in the \emph{first} layer. Since the
first stage of a network processes non-binary data, BinaryNet does not
feature binary optimization for this layer. However if the input data
is split into its constituent bit-planes, binary optimization can
still be applied. In particular, we split the input vector in a matrix
of $8$ rows, and recombine the result after multiplication by a
weighted sum. Our experimental results report an overall
\speedup{\approx 3} performance boost when comparing the full binary
optimized network with one in which the first layer is not binary
optimized.

Finally, in terms of memory the \gpustar{} implementation requires
\MB{4.57} instead of \MB{140.6} as in the case of non binary optimized
implementation, resulting in a saving \speedup{\approx 31} amount of
memory.

\subsection{Convolutional Neural Network on CIFAR-10}
\label{sec:cnncifar}

\begin{table}[htbp]
  \centering
  \caption{Average prediction time of the BCNN.}
  \label{tab:res-cnn}
  \begin{tabular}{rrr}
    \toprule
     Espresso \cpu{} & Espresso \gpu{} & Espresso \gpustar{} \\
    \midrule
    \ms{85.2} & \ms{5.2} (\speedup{16}) & \ms{1.0} (\speedup{85}) \\
    \bottomrule
  \end{tabular}

\end{table}

To the best of our knowledge, no BDNN implementation of
\emph{binary-optimized} CNN layers is publicly available. Our
self-evaluation implements the \emph{VGGNet}-like CNN architecture
from Hubara et al.~\cite[Sec. 2.3]{hubara2016binarized}, and evaluates
it across our three modalities: as expected the $\gpustar$
implementation achieves significantly better performance.

\paragraph{Unrolling and pooling}

Note how the \gpu{} implementation offers a slightly better
improvement over \cpu{} with respect to the MLP test, with an
\speedup{\approx 16} speed-up. In this experiment, the inherent
parallelism of unrolling and pooling, and the GPU higher memory
throughput explain the behavior. Gains are marginal as FMA still
represents the computational bottleneck.

\paragraph{Bit-packing}

The \gpustar{} implementation results in a \speedup{\approx 5}
performance gain with the respect to \gpu{}. These gains, to binary
optimizations, are slightly smaller than those discussed for MLP
in \Section{mlpnist}. The output of convolutional layers is
significantly larger than those of MLP's dense layers, therefore, the
computation of bit-packing sign-activation requires more computational
effort.

Finally, in terms of memory the \gpustar{} implementation requires
\MB{1.73} instead of \MB{53.54} as in the case of non binary optimized
implementation, resulting in a saving \speedup{\approx 31} amount of
memory.

\section{Conclusions}
\label{sec:conclusion}

In this paper we presented \espresso{}, a highly optimized
forward-propagation framework for both traditional DNNs as well as
BCNNs, that supports heterogeneous deployment on CPU and GPU. While
BinaryNet and Nervana/neon BDNN implementations are limited to MLP
networks, our framework also supports the popular CNN while
simultaneously outperforming state-of-the-art implementations of MLP
networks. \espresso{} is highly-efficient, light-weight and
self-contained. Computation on the GPU side is done though
specifically designed CUDA kernels, which, combined with a more
careful handling of memory allocation and bit-packing, allows us to
obtain considerable performance improvements. In future work we would
like to add training capabilities, and perform additional performance
comparisons on larger standard datasets.

\bibliographystyle{iclr2018_conference}
\bibliography{ref}

\end{document}